# DCELANM-Net:Medical Image Segmentation based on Dual Channel Efficient Layer Aggregation Network with Learner

Chengzhun Lu[1], Zhangrun Xia[1], Krzysztof Przystupa[2], Orest Kochan[1,3], Jun Su[1]

Correspondence
Chengzhun Lu, School of Computer Science
Hubei University of Technology, Wuhan, Hubei, China
E-mail :lcz529959@126.com

ACKNOWLEDGMENTS
Throughout the writing of this dissertation I have received a great deal of support and assistance.
I would first like to thank my supervisor, Jun Su, whose expertise was invaluable in formulating the research questions and methodology. Your insightful feedback pushed me to sharpen my thinking and brought my work to a higher level.
I would particularly like to acknowledge my teammate/group mate/team members, Zhangrun Xia, Krzysztof Przystupa and Orest Kochan for their wonderful collaboration and patient support

**Abstract.** Segmenting medical images is a principal component of computer vision. The UNet model framework has taken over as the standard framework for this activity across a wide range of medical picture segmentation applications. Due to CNN's convolution operation limitations, the model's global modeling ability isn't absolutely perfect. Moreover, a single convolution operation cannot gather feature information at various scales, which will have an impact on the quality of the global feature extraction as well as the localization of local details. The DCELANM-Net structure, which this article offers, is a model that ingeniously combines a Dual Channel Efficient Layer Aggregation Network (DCELAN) and a Micro Masked Autoencoder (Micro-MAE). On the one hand, for the DCELAN, the features are more effectively fitted by deepening the network structure; the deeper network can successfully learn and fuse the features, which can more accurately locate the local feature information; and the utilization of each layer of channels is more effectively improved by widening

[1] School of Computer Science, Hubei University of Technology, Wuhan , 430068, China

[2] Department of Automation, Lublin University of Technology, Nadbystrzycka 36, 20-618 Lublin, Poland

[3] Department of Measuring Information Technologies, Lviv Polytechnic National University, Bandery Str. 12, 79013 Lviv, Ukraine

the network structure and residual connections. We adopted Micro-MAE as the learner of the model. In addition to being straightforward in its methodology, it also offers a self-supervised learning method, which has the benefit of being incredibly scaleable for the model. This scalable method enables the generalization of high-volume models, and the models can show good scaling behavior. Using only around 10% of UNet's parameters, we show that the suggested DCELAN can virtually replace UNet while enhancing performance by about 9.6%. We also show that Micro-MAE is a powerful and adaptable learner that we can incorporate into our network design and give the model good accuracy and stability for tasks involving medical picture segmentation. Superior metrics and good generalization are obtained by DCELANM-Net on the datasets Kvasir-SEG and CVC-ClinicDB.

**KEYWORDS**
medical image segmentation, CNN, self supervised learning, transformer

# 1 Introduction

Convolutional neural networks (CNN) simplify computer vision tasks like picture segmentation. Full convolutional networks (FCNs) [1] revolutionized picture segmentation. FCN [1] substitutes the fully connected layer behind the classic CNN with the convolutional layer to output the thermal map instead of the category. Upsampling recovers image size to segment medical images. UNet [2], a symmetric encoder-decoder and jump connection architecture, excels in medical picture segmentation. This structure segments well with a few data sets. Thus, advances in U-shaped architecture are increasing. Inspired by ResNet, Res-UNet individually built a jump connection for each submodule in UNet [2,3,4], increasing network depth, preventing overfitting, and improving model correctness. For dense connectivity of UNet submodules, DensNet is used by Dense-UNet [2,5,6]. The end-to-end dense connection network makes the network easier to train and is conducive to the backward propagation of gradient. These outcomes mostly act on UNet [2] submodules.

Inside the submodule, there are also operations to improve the CNN structure. The backbone network, for example, employs image classification networks such as VGG, GoogLeNet, and ResNet [4,7,8]. ResUNet++ is an improved version of ResUNet [3,9]. It combines ResNet and UNet, introducing squeeze and excitation blocks, SENet attention blocks, and the DeepLabv2 ASPP module [2,4,10,11]. It has its own section. Furthermore, MultiResUNet is a similar enhancement. In UNet [2,8,12], the MutiRes block (similar to the inception block in GoogLeNet) was used to replace the traditional convolution module. The MutiRes block replaced the two $3\times3$ convolutions in UNet with parallel and merged $3\times3$, $5\times5$, and $7\times7$ convolution operations. In UNet, the skip connection is replaced by a ResPath connection [2,3]. This has the advantage of allowing the encoder's low-level features to reach a consistent depth with the decoder's high-level features before concatenation. DCUNet [13] has the same thought. It replaces the MultiRes block in MultiResUNet [12] with a dual-channel (DC) block and improves segmentation accuracy by using dual-channel and multi-scale data [12,13]. In addition, to reduce the number of model parameters, all 5×5 and 7×7 convolution operations in MultiResUNet are replaced with two 3×3 convolution operations. As a result, while reducing model parameters, DCUNet maintains segmentation accuracy [12,13].

In addition, with the emergence of self-attention mechanisms, particularly the transformer [14] architecture, Transformer was first proposed to address the problem of machine translation, and it has since become the predominant framework for natural language processing (NLP). It can pre-train on some larger datasets and then fine-tune on smaller datasets [15]. Many computer vision tasks have supplanted CNN architectures with transformer architectures, inspired by the success of NLP [14,16]. With the advancement of technology, Vision Transformer [17] (ViT) has excelled in the field of computer vision and performed admirably on numerous duties. Therefore, segmentation of medical images based on a self-attention mechanism is extensively utilized. AttentionUNet, for example, proposed an attention gate that is plug-and-play and can be directly inserted into the UNet model to suppress irrelevant regions in the input image and emphasize the salient features of particular local regions [2,18]. At the skip connection, the high- and low-level features of the decoder and encoder are exhaustively utilized. In addition to UNet as the overall network architecture, the encoder structure uses TransUNet of transformer [2,14,19]. The local features extracted in CNN are combined because the self-attention mechanism in transformer [14] has the advantage of extracting global features. Segmentation of medical images produces favorable results.

Masked signal modeling (MSM) [20] is a popular self-supervised pre-training job that removes part of the input signal and predicts it. Language, vision, and speech use this task. Recently, masked image modeling (MIM) [21] has become a strong contender for supervised pretraining methods widely used in computer vision, with MIM-based pretrained models achieving very high fine-tuning accuracy on a wide range of vision tasks of different types and complexity. After that, masked autoencoder (MAE) successfully pretrained a ViT for image analysis [17,22]. Mask random regions of the input image to recreate missing pixels.

In summary, we develop Micro-MAE with an asymmetric encoder-decoder structure on MAE[22] as a learner to extract global features to increase the accuracy of medical image segmentation, consequently enhancing the accuracy of medical image segmentation. We propose DCELANM-Net as the backbone network block. To sum up, the following is a list of our major contributions:

We propose the DCELANM-Net framework. Several layers interchange feature information across scales by using DCELAN PATH as a jump connection. DCELANM-Net makes use of DCELAN Blocks to successfully fuse this feature information in order to deepen and widen the model while gaining improvement in accuracy and fewer parameters.

The Micro-MAE is used as a learner for up- and downsampling to obtain contextual feature information more effectively, and the MAE [22] is used as an encoder to extract global features to improve the accuracy of segmentation of medical images. This optimizes the accuracy of the whole model while reducing memory consumption in the MAE [22] and ELAN Block structures, which benefits an improvement in the overall accuracy of the model.

Testing the performance of DCELANM-Net on the challenging Kvasir-SEG and CVC-ClinicDB [23,24] datasets with fewer parameters and better performance when compared to previous work. Additionally, performance metrics on another class of medical image datasets are also significantly better than the existing state-of-the-art models.

# 2 Related Works

Combining neural networks and Micro-MAE Since the semantics of colorectal cancer tumor images are relatively simple and the structure is relatively fixed, both high-level semantic information and low-level features are very essential, so we use the U-shaped structure based on the encoder and decoder as the main architecture of the model. We designed a novel backbone network, the DCELAN Block. It is a deformable component based on ELAN [25] in the YOLOv7 network structure. Originally introduced by VoVNet and CSPVoVNet, in addition to contemplating the architecture for extracting features, it also analyzes the gradient path to effectively learn and fuse by controlling the shortest, longest gradient path while expanding the breadth and depth of the network [26,27]. In order to acquire multi-scale information, we take inspiration from the structure of DCUNet and change the ELAN Block into DCELAN to surmount the problem of insufficient spatial features [14,25]. Moreover, residual structure is added to make the parameters pass effectively in a deeper and wider network. We replace the Res-Path skip connection structure in MultiResUNet [13] and use DCELAN PATH. The DCELAN PATH Block enables the DCELAN Block to have the ability to concatenate features learned from images at various scales. We also use ViT as a pre-trained model, and since there is no ImageNet analogous dataset for medical image segmentation, we also conduct self-pre-training on the same dataset for downstream tasks [17,28,29]. In this case, MAE [22] is embedded at the juncture of downsampling and upsampling as a scalable self-supervised learner, which arbitrarily masks off part of the patches and reconstructs the missing pixels. Finally, the reconstructed image was input into the network. So far, DCELANM-Net has been implemented, which has been very successful in medical image segmentation.

# 3 Method

## 3.1 DCELANM-Net

In this part, the comprehensive model architecture of DCELANM-Net, shown here in Figure 1, is presented. When the image has been processed, it is loaded into

DCELANM-Net, where it undergoes the following steps: First, the features are retrieved, and then the feature map is reduced using four DCELAN Blocks and four downsamplings. Following this, the feature information that was extracted by the fourth DCELAN Block is input into Embedded Micro-MAE. The feature information is then passed into the encoder by masking a portion of the feature information. Finally, the feature information of the encoder is learned and passed onto the decoder so that it can reconstruct the feature information. Upon the completion of the reshaping procedure, the information regarding the reconstructed features is transmitted to the backbone network. At this time, the feature information that was obtained from the left half of the DCELAN Block and the downsampled feature information is first catalyzed with the feature information that was obtained from the right half of the DCELAN Block and the upsampled feature information, and then the residuals are connected. This is done by cascading paths and residuals. Four DCELAN Block, upsampling, and DCELAN PATH splicing operations are performed on the entering feature map to accomplish the same thing. The product of the output process in its entirety is acquired.

Motivated by Res-Path [13] Path's work. In order to construct the DCELAN PATH structure, we feed the encoder and decoder information into a convolutional block that consists of CBS.1 and CBS.3. The CBS Block is detailed in the reference of the first chapter (1) in 3.2 Backbone Block. The ability to concatenate characteristics that have been learned from images of different scales is made possible by the DCELAN PATH Block, which is part of the DCELAN Block. CBS.2 is used to extract features, CBS.1 is used as a residual join, and the convolution kernel is set to $1 \times 1$ so that the dimensionality can be reduced even more. This helps reduce the amount of processing and memory that is required. It simplifies the process of learning and has been demonstrated to have significant potential in the field of medical picture analysis [29]. By cascading the encoder and the decoder with the features, it is possible to prevent the loss of some feature information.

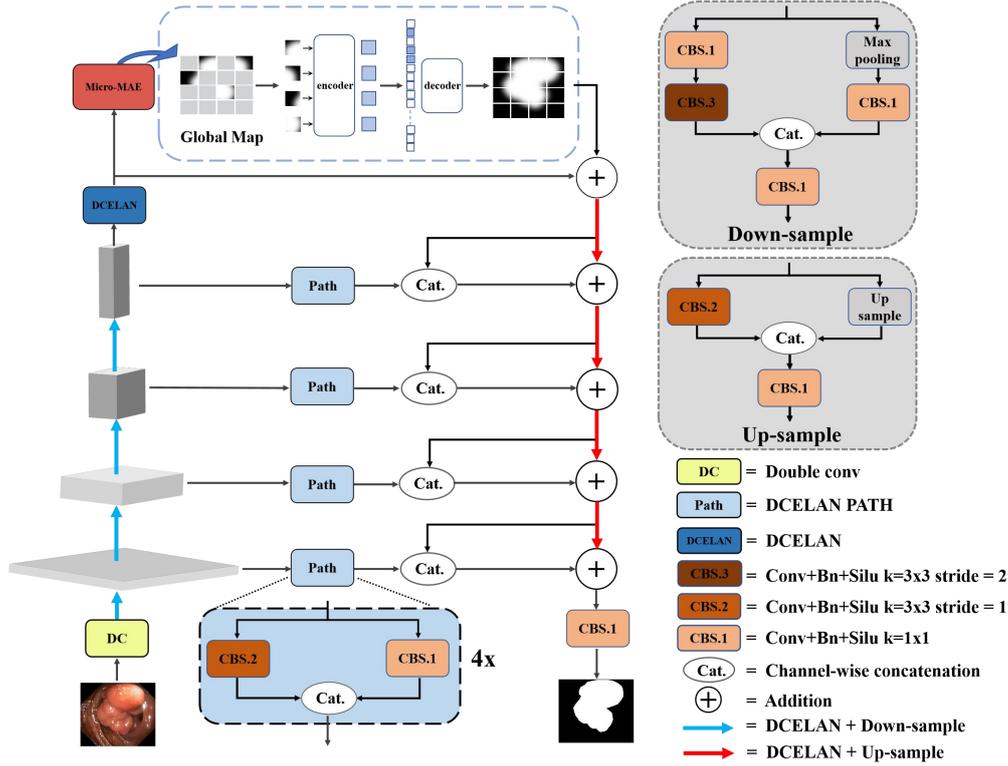

Figure 1: DCELANM-Net framework map, which introduces Embedded Micro-MAE, downsamples feature information combined with DCELAN PATH and outputs a combination of feature maps to generate a global map. For more information, see Section 3.1.

## 3.2 Backbone Block

### (1) Dual Channel Efficient Layer Aggregation Networks

When it comes to the process of creating effective network structures, it is essential to take into account not only the level of accuracy that the model provides but also the number of parameters and the amount of computational work that the model requires. The development of EfficientNet [30] demonstrated conclusively that the depth, width, and resolution of a CNN are the most important factors to consider when attempting to increase the performance of such a network. As a result, we combined the concepts of scaling the channels and merging them. In turn, we designed a balanced integrated network, which is depicted here as Figure 2: DCELAN Block and ELAN Block. DCELAN Block is a composite scaling search that is applied on top of ELAN Block. It is determined by the depth, width, and input size. We define three CBS (Convolutional + BatchNormal + Silu) Blocks, CBS.3 convolution kernel $3 \times 3$ with step size 2, CBS.2 convolution kernel $3 \times 3$ with step size 1, and CBS.1 convolution kernel $1 \times 1$ with step size 1. Use CBS.1 for channel downscaling and deepening the model depth when the image input comes in. After that, copy one side to spread the model width into a double channel. Next, we use 8 CBS.2 for feature extraction, and following that, we connect a residual block to the output of these three blocks so that you may fuse them together to create stitching channels. In the end, employing CBS.1 brings the number of channels down to the same level as the number of channels present in the original input. Although ELAN only has one channel, the only thing that changes between the module and the DCELAN Block is the splicing of the residual block CBS.1 with the feature information of the output on the left. Finally, the number of channels is still reduced to the number of channels that were present in the original input by CBS.1.

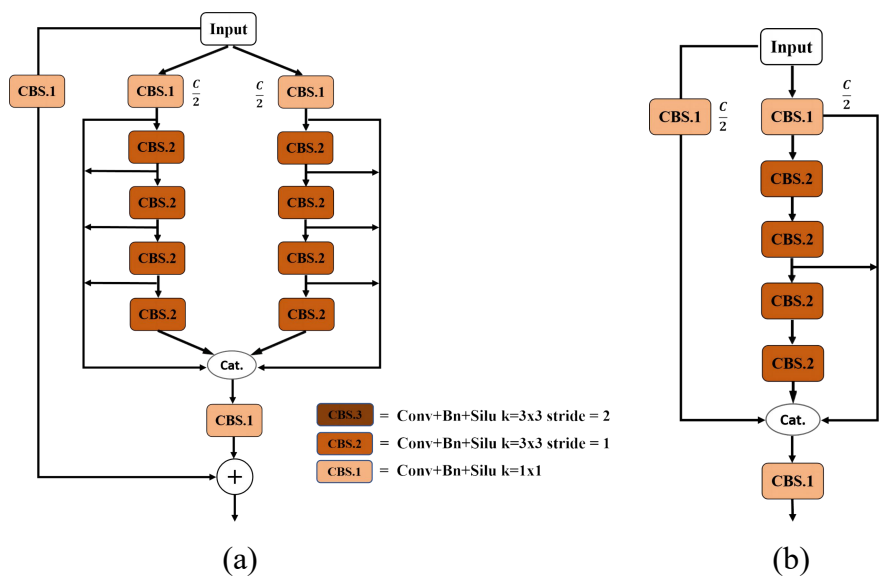

Figure 2: The (a) on the left indicates DCELAN Block, and the (b) on the right indicates ELAN Block

## (2) DCELAN Up sampling and down sampling

We also combine the concepts of channel scaling and merging to address the issues of feature information loss during downsampling as well as picture edge blurring and low calculation accuracy caused by upsampling. In downsampling, we employ CBS.3 with step size 2 in addition to maxpooling with step size 2 to minimize the dimension of the channels before merging them. By doing this, the CBS.3 feature information extraction and the feature information lost during maxpooling may be combined to prevent feature information loss. Similar to the downsampling, the channels' dimensions are decreased and then merged using CBS.2 with a step size of 2, in addition to bilinear interpolation. This increases computation accuracy while also reducing image edge blur. It is simpler to aggregate characteristics at skip links across the model thanks to these two samplers.

## 3.3 Embedded Micro-MAE

MAE is based on two fundamental designs, the first of which is an asymmetric architecture. Randomly visible sections of a picture are fed into an encoder in this design, and the original image is rebuilt in another decoder. The second goal is to attain the highest level of precision feasible by contrasting large-scale trials with the masked input image. In medical segmentation, we insert embedded Micro-MAE modified on MAE between upsampling and downsampling, take the highest feature map as input, scale the masked image to 75%, and stretch the unmasked chunks into vectors to do ViT [17] in the encoder, and the decoder will reconstruct all the image information in the masked parts back, which will recover the feature information lost due to the multi-layer downsampling. As seen in Figure 3.

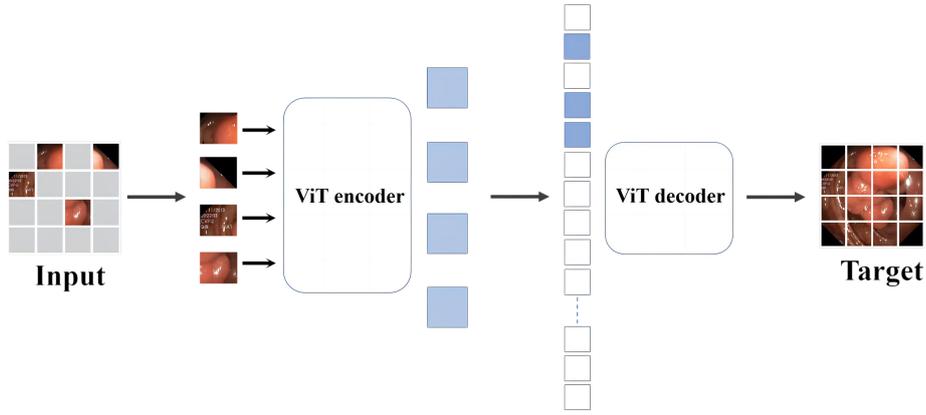

Figure 3: Embedded Micro-MAE A model for an asymmetric encoder-decoder architecture

## (1) Patch Embedding

The high-dimensional data must first be transformed into sequences before the embedding layer can continue. This is necessary because the ViT [17] algorithm accepts only sequences as input. In the processing of 2D images, making a 2D body requires that it be first shaped into a series of flattened two-dimensional surfaces, where C is an input channel, $N=HW/P^2$ is the number of face pieces, (H,W) is the input resolution, and (P,P) is the face piece resolution, i.e., the length of the input sequence entered into the transformer. After that, a trainable linear projection is used to transfer them to the faceted embedding, $\mathbf{E} \in \mathbb{R}^{(P^2 \times C) \times D}$, which is a face sheet embedded projection, $\mathbf{E}_{pos} \in \mathbb{R}^{(N+1) \times D}$ and is position embedded, respectively. Add the patch sequence x embedding along with the position sequence embedding to keep track of the location. The state at the output of the transformer encoder ($z_L^0$) is utilized as the picture representation in the standard ViT [17], which employs 1D learnable position embedding (Equation 4). However, it was discovered through experimentation [31] that the learnable 1D positional embedding had an impact on the Micro-reconstruction MAE's capabilities. As a result, in the pre-training stage, we employ the sine - cosine form of the location embedding. We keep working with the learnable position embedding approach for the downstream task, but we initialize it with sine-cosine embedding values.

$$z_0 = [x_{class}; x_p^1 E; x_p^2 E; \cdots; x_p^N E] + E_{pos} \quad E \in \mathbb{R}^{(P^2 \times C) \times D}, \quad E_{pos} \in \mathbb{R}^{(N+1) \times D} \quad (1)$$

Transformer block consists of alternating layers of Multihead Self-Attention (MSA) [32] and Multilayer Perceptron (MLP) block, consisting of LN representation, layer normalization operator, and $z_L$ encoded image representation.

$$z'_\ell = \text{MSA}(\text{LN}(z_{\ell-1})) + z_{\ell-1} \quad \ell = 1...L \quad (2)$$

$$z_\ell = \text{MLP}(\text{LN}(z'_\ell)) + z'_\ell \quad \ell = 1...L \quad (3)$$

$$y = \text{LN}(z_L^0) \quad (4)$$

## (2) Encoder and Decoder in Micro-MAE

Micro-MAE is an asymmetric encoder-decoder structure made up of ViT [17], but it can only be used on visible, unmasked patches. The encoder embeds patches by adding linear projections of position embeddings, just like in normal ViT [17], and then processes the resulting set using a sequence of transformer blocks. However, the encoder only uses a small portion of the entire set-let's say 25%-for operation. Patches that were blocked are unblocked; no mask token is used. the capacity to train extremely large encoders with a small amount of computation and memory. The entire collection of tokens, including the encoded visible patch and the mask token, is the input to the micro MAE decoder. Every mask token is a collectively learned vector that denotes the existence of missing patches that need to be anticipated. Include position embeddings for each token in this entire set; otherwise, the mask tokens wouldn't know where they were in the image. Another series of transformer modules is present in the decoder.

## 3.4 Loss function

By anticipating pixel values for each mask patch, Micro-MAE reconstructs the input. The loss function computes the mean squared error (MSE) in pixel space between the reconstructed picture and the original image. Tversky Loss is the loss function that applies to the entire network. This can be found in (6). The Tversky coefficient is a generalization of the Dice coefficient and the Jaccard coefficient, and it is utilized to find a solution to the problem of data imbalance and achieve a better equilibrium between accuracy and recall. The Tversky coefficient is equivalent to the Dice coefficient when and are both set at $\alpha = \beta = 0.5$. When $\alpha = \beta = 1$, the Tversky coefficient is the Jaccard coefficient. By manipulating the values of $\alpha$ and $\beta$, it is possible to exercise control over the ratio of false positives to false negatives. $B$ represents the actual mask, while $A$ represents the one that was predicted. The result of $|A-B|$ is a false positive, and the result of $|B-A|$ is a false negative. The following is the formula for MSE Loss, as well as the formula for Tversky Loss:

$$loss(x_i, x_j) = (x_i - x_j)^2 \quad (5)$$

$$T(A,B) = \frac{|A \cap B|}{|A \cap B| + \alpha |A-B| + \beta |B-A|} \quad (6)$$

Our loss function is denoted as $\mathcal{L} = x\mathcal{L}_{TL}^w + z\mathcal{L}_{MSE}^w$, where $\mathcal{L}_{TL}^w$ and $\mathcal{L}_{MSE}^w$ represent the weighted IOU Loss and MSE Loss, respectively, for global and local (pixel-level) limitations, respectively. When compared to the ordinary IOU loss, which is commonly used in segmentation tasks, the weighted IOU loss increases the weights of hard pixels to underline their importance. Furthermore, rather of allocating the same weight to each pixel, the usual MSE loss method prioritizes hard pixels. Their utility in the realm of dominating object identification has been established [33,34]. . The following is how we decided on the loss coefficients of x = 0.8 and z = 0.2.

$$\mathcal{L} = 0.8 \times \mathcal{L}_{TL}^w + 0.2 \times \mathcal{L}_{MSE}^w \quad (7)$$

# 4 Experiments and Discussion

## 4.1 Datasets and Evaluation

To evaluate DCELANM-Net's performance on medical datasets, we selected two public datasets for adenoma colonoscopy detection and a large public dataset for dermoscopy. Table 1 describes succinctly the public datasets required for the experiment.

Kvasir-SEG [23] is the first multi-class gastrointestinal illness dataset. The original Kvasir dataset consists of 8,000 GI track photos from eight unique categories, each with 1,000 images. Kvasir-SEG segmented polyp dataset updated 13 polyp class photos with new images to improve dataset quality.

CVC-ClinicDB [24]: an open dataset containing 612 pictures from 31 colonoscopy sequences at 384 284 pixels. It detects colonoscopy video adenomas in medical image segmentation. It is a library of pictures derived from recordings of colonoscopies. The collection includes polyp frames and ground truths. The ground truth picture consists of a mask that matches to the region of the image covered by the polyp.

ISIC2018 [35]: ISIC 2018 provides for the first time a separate dataset for the task, containing 2594 dermoscopic raw images and 2594 corresponding binary label images, which contain skin diseases including photochemical keratosis and intraepithelial neoplasia, benign keratosis, basal cell carcinoma, squamous cell carcinoma, dermatofibroma, melanoma, nevi, and vascular lesions.

Our experimental results will utilize the measurement metrics dice coefficient (mDice), mean intersection over union (mIOU), and mean precision (mPre). They are commonly employed for medical image segmentation tasks.

Table 1:Overview of the Datasets.

| Modality | Dataset | NO. of images | Original resolution |
|---|---|---|---|
| Enteroscope | Kvasir-SEG [23] | 1000 | Variable |
| Endoscopy | CVC-ClinicDB [24] | 612 | 384x288 |
| Dermoscopy | ISIC2018 [35] | 2594 | Variable |

## 4.2 Implementation Details

Pytorch was used in the experiments to construct the network model, which was written in Python 3 [36]. Experiments were carried out using a desktop computer that featured an Intel Core i7-12700H processor clocked at 2.7 GHz, 16.0 GB of RAM, and an NVIDIA GeForce RTX 3060 (GPU).

For the purpose of colonoscopy image segmentation, we self-trained Micro-MAE on a single NVIDIA Tesla V100 GPU, employing the Table 1 dataset. The model was trained for 500 rounds using cosine annealing with an 8-batch size and 1e-4 learning rate. It took roughly 20 hours to train the model for this many rounds, and then we used the Adam optimization approach to optimize the overall parameters. Zero-filled squares were increased to $256 \times 256$ for training images. This preserved the image's aspect ratio. We also used multiple scales. Randomly resize the picture to a multiple of 32 between 192 ($256 \times 0.75$) and 320 ($256 \times 1.25$). Comparing the baseline model UNet [2], which is provided as [16,32,64,128,256] for the encoder filter and [256,128,64,16] for the decoder filter, is the next step. Similar to Table 2. To keep channels in their original state, the backbone network block performs dimensionality reduction splicing after the model extracts features. The self-trained Micro-MAE receives the reshaped data when the channel size is 256. The encoder encodes the location in Micro-MAE, the feature map is split into $4 \times 4$ patches, 75% of the feature information is randomly masked, and the reconstructed feature information is acquired. "Feature reconstruction" describes this. The segmentation effect is achieved after upsampling, cascading, and creating the DCELAN PATH link to the encoder and decoder.

Table 2: Details of DCELANM-Net

| DCELANM-Net | | | |
|---|---|---|---|
| Block | Filters | Path | Filters |
| DCELAN Block 1<br>DCELAN Block 9 | 16 | DCELAN PATH 1 | 16×8 |
| DCELAN Block 2<br>DCELAN Block 8 | 32 | DCELAN PATH 2 | 32×6 |
| DCELAN Block 3<br>DCELAN Block 7 | 64 | DCELAN PATH 3 | 64×4 |
| DCELAN Block 4<br>DCELAN Block 6 | 128 | DCELAN PATH 4 | 128×2 |
| DCELAN Block 5 | 256 | NULL | NULL |

## 4.3 Result

### 4.3.1 Parameter

In order to determine the efficacy of DCELANM-Net, we devised tests on three different medical datasets. The results of these experiments are presented in Table 3, along with the number of parameter metrics and the number of floating point operations necessary to apply the model. There are also models that are based on CNN that can be used (called transformer models), such as DCSAU-Net, PraNet, and HarDNet-MSEG [37,38,39]. Although our DCELANM-Net is more complex than the majority of CNN-based neural networks, it is nevertheless more effective in terms of GFlops than other transformer-based techniques. For all of the comparison models, the size of the image input is 256×256.

Table 3: Parameters(Million) and GFlops

| Model | Type | Parameters(Million) | GFlops |
|---|---|---|---|
| U-Net (baseline) [2] | CNN | 43.93 | 43.47 |
| DCSAU-Net [35] | CNN | 34.51 | 35.15 |
| MultiResUNet [12] | CNN | 29,06 | - |
| DCUNet [13] | CNN | 10,07 | 21.5 |
| Colonformer [42] | Transformer | 52.94 | 22.94 |
| TransUNet [19] | Transformer | 105.28 | 60.75 |
| PraNet [36] | Transformer | 32.55 | 13.11 |
| **DCELAN** | **CNN** | **8.18** | **33.43** |
| **DCELANM-Net** | **Transformer** | **28.35** | **40.24** |

There is a correlation between the size of the convolution kernel and the number of channels of input and output when discussing the number of parameters. After performing the dimensionality reduction computation, each channel in our DCELAN Block is then stitched together; to put it another way, the convolution operations are carried out in low dimensions. And all of the convolutional operations are carried out with the help of the CBS algorithm. And in the Micro-MAE encoder, the encoder does not fill with 0 for the matched patches that are masked; rather, it ignores them directly. Because of this, reducing the number of patches can directly reduce the computation, speed up the training by three times the rate of 75% of the mask rate, and reduce the amount of memory consumption, which enables larger models to be used. In addition to that, it possesses a two-fold multi-resolution feature, which explains why the outcomes are superior to the model that was used for comparison.

DCELAN has a modest cost and only 8.18M parameters, but it can achieve up to 33.43 GFlops. So, in order to further strengthen the GFlops and make the model more robust, we made the decision to sacrifice some parameters. That is to say, DCELANM-Net is successful in achieving GFlops with 28.35M parameters.

### 4.3.2 Ablation Experiments

The following ablation experiments were carried out by us in order to evaluate the effectiveness of the DCELANM-Net framework in a variety of diverse environments. The results are shown in Table 4.

In our studies, we compared the performances of ELAN, DCELAN, and DCELAN + Micro-MAE on two different datasets. These datasets were CVC-ClinicDB and Kvasir-SEG [23,24], respectively. As compared to ELAN, the results of the studies show that using DCELAN results in improvements in mDice and mIOU of 4.82% and 2.78%, respectively, on the CVC-ClinicDB [24] dataset.

When compared to ELAN, the performance of mDice and mIOU under DCELAN+Micro-MAE was enhanced by 6.77% and 5.37%, respectively. DCELAN performed significantly better than ELAN, mDice, and mIOU on the Kvasir-SEG [23] dataset, with a 6.03% and 5.43% improvement, respectively. When compared to ELAN, mDice, and mIOU, the performance of DCELAN+Micro-MAE was enhanced by 7.90% and 5.78%, respectively. As a result, it is possible to demonstrate that the performance of the combined DCELANM-Net framework is better when compared.

Table 4: Ablation Experiments

| datasets | Methods | mDice | mIOU | mPre |
|---|---|---|---|---|
| CVC-ClinicDB [24] | ELAN | 0.8365 | 0.8154 | 0.8612 |
|  | DCELAN | 0.8847 | 0.8432 | 0.9014 |
|  | **DCELAN+MAE** | **0.9042** | **0.8691** | **0.9322** |
| Kvasir-SEG [23] | ELAN | 0.8122 | 0.7858 | 0.8029 |
|  | DCELAN | 0.8725 | 0.8401 | 0.9085 |
|  | **DCELAN+MAE** | **0.8912** | **0.8436** | **0.9218** |

### 4.3.3 Comparison with others

Early diagnosis throughout the progression of polyps from small to large before they turn into colorectal cancer can minimize the lesion rate and enhance patient survival. Polyps are often prominent and have an uneven shape. In this article, we still compared our results to those obtained from the original UNet benchmark model. Also, we employed the same experimental setup with the selected CVC-ClinicDB and Kvasir-SEG [23,24] datasets.

Table 5 displays the segmentation results in CVC-Clinic DB [24], which still has an improvement over other State-of-the-Art(SOTA) networks. Furthermore, the visualization results show that DCELANM-Net has better segmentation results in some challenging images (such as polyps), which are commonly missed in colonoscopies.

The results of the Kvasir-SEG [23] segmentation are displayed in Table 5, and we discovered that DCELANM-Net improves mDice by 10.91% and mIOU by 2.95% in comparison to UNet. In general, DCELANM-Net has higher segmentation performance for medical images of the same lesion when using multiple datasets. This clearly demonstrates the model's robustness.

Table 5: Comparison with others

| Dataset | CVC-ClinicDB[24] | | | Kvasir-SEG[23] | | |
|---|---|---|---|---|---|---|
| Mertric | mDice | mIOU | mPre | mDice | mIOU/JS | mPre |
| UNet (baseline) [2] | 0.8464 | 0.7730 | 0.8496 | 0.7821 | 0.8141 | 0.7241 |
| UNet++ [40] | 0.794 | 0.8115 | 0.9192 | 0.8210 | 0.743 | 0.8512 |
| ResUNet++ [9] | 0.8321 | 0.7811 | 0.8705 | 0.8074 | 0.7231 | 0.8991 |
| DCUNet [13] | 0.765 | 0.8084 | 0.9261 | 0.5018 | 0.6385 | 0.8606 |
| DCSAU-Net [35] | 0.8816 | 0.8520 | 0.9093 | 0.8569 | 0.7842 | 0.8850 |
| DoubleU-Net [41] | **0.9239** | 0.8366 | 0.8923 | 0.8620 | 0.8021 | 0.8721 |
| Attention-UNet [18] | 0.8655 | 0.8358 | 0.9046 | 0.8124 | 0.7954 | 0.8630 |
| TransUNet [19] | 0.8411 | 0.7991 | 0.8763 | 0.8213 | 0.7855 | 0.8568 |
| PraNet [36] | 0.8999 | 0.8495 | 0.9218 | 0.887 | 0.8402 | 0.9153 |
| **Ours** | 0.9042 | **0.8691** | **0.9322** | **0.8912** | **0.8436** | **0.9218** |

## 4.3.4 Generalization to Other Datasets

In the ISIC2018 [35] dataset, an automated skin lesion diagnostic instrument scored higher for its ability to detect melanoma, thereby increasing survival rates in many cases. In the experiment to validate DCELANM-Net's ability to generalize, we further evaluated our model using ISIC2018 data. We discover that our model is also capable of achieving more optimal evaluation metrics with larger datasets. Our method obtains mDice of 0.912, mIou of 0.8912, and mPre of 0.9231 according to Table 6 of ISIC2018 [35]. Compared to other SOTA networks, we can observe a greater improvement. In the experiment conducted to validate DCELANM-Net's generalization capability, the outcomes demonstrate the superior performance and learning ability of DCELANM-Net.

Table 6: Generalization to Other Datasets

| Dataset | ISIC2018[35] | | |
|---|---|---|---|
| Mertric | mDice | mIOU | mPre |
| UNet (baseline) [2] | 0.8554 | 0.8025 | 0.8837 |
| UNet++ [40] | 0.8094 | 0.8122 | 0.8997 |
| ResUNet++ [19] | 0.8557 | 0.8100 | 0.8705 |
| DCUNet [13] | 0.6556 | 0.8173 | 0.8523 |
| DCSAU-Net [35] | 0.8526 | 0.8301 | 0.9151 |
| DoubleU-Net [41] | 0.8962 | 0.8136 | 0.9039 |
| Attention-UNet [18] | 0.8547 | 0.8149 | 0.9153 |
| TransUNet [19] | 0.8186 | 0.7700 | 0.8473 |
| PraNet [36] | 0.9024 | 0.8906 | 0.9145 |
| **Ours** | **0.9126** | **0.8912** | **0.9231** |

### 4.3.5 Visualization

Our model was tested using public datasets and compared to several other approaches. Figure 4 shows that our model beats competitors on multiple datasets with good applicability and robustness. DCELANM-Net segmentation findings on various test sets indicate that our model recognizes polyp tissue size, homogeneous areas, and varied textures. Our model's image segmentation maps are virtually identical to the real mask (GT) in microscope photos, indicating that it can properly locate and segment polyp tissue in many challenging conditions. Our model performs well in medical picture segmentation tasks and may be used for medical diagnosis and therapy. Our discoveries can help physicians better diagnose and treat disorders. Our method also works for medical image analysis applications such as picture alignment, target detection, and classification. Our research suggests new paths for medical image processing, which can help individuals understand and treat many medical diseases.

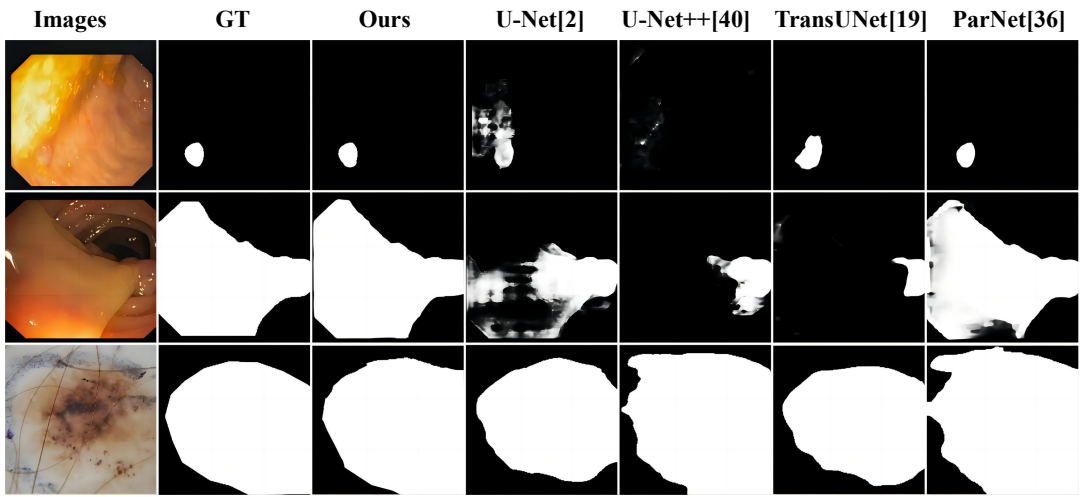

Figure 4: Each row from top to bottom reflects the experimental operations on the CVC Clinical DB, Kvasir SEG, and ISIC2018 [23,24,35] datasets. Columns provide the visualisation results of various models.

# 5 Conclusion

In this paper, we propose a medical image segmentation model called DCELANM-Net, which combines a two-channel residual aggregation network with a micro mask self-encoder, and enhances the existing SOTA classification and segmentation performance by this combination. In addition, a method dubbed Micro-MAE is developed in this study for adapting to embedded medical picture segmentation tasks with good results. The experimental results in this work reveal that DCELANM-Net has good performance metrics on both CVC-ClinicDB and Kvasir-SEG [23,24] datasets, and can effectively increase the accuracy of automatic segmentation of polyp-detecting lesions from colonoscopy pictures. Moreover, the performance metrics of DCELANM-Net on the ISIC2018 [35] dataset are much superior than the existing state-of-the-art models, which highlights the huge potential of DCELANM-Net in the field of medical image segmentation. Yet, the generalization ability is equally necessary for a good medical picture segmentation model. Thus, in future work, we will study the feasibility of employing the DCELANM-Net model for other large datasets to improve the performance metrics of the model. We hope that the growth of the dataset can overcome picture duplicate information more efficiently and give improved support for clinical diagnosis and therapy.

**ORCID**
Chengzhun Lu: https://orcid.org/0009-0007-8397-5784

Zhangrun Xia: https://orcid.org/0009-0002-6691-9119
Jun Su: https://orcid.org/0000-0002-4290-5049
Orest Kochan: https://orcid.org/0000-0002-3164-3821
Krzysztof Przystupa: https://orcid.org/0000-0003-4361-2763